\documentstyle[12pt,psfig]{article}
\begin{document}
\def\refitem{\par\parskip 0pt\noindent\hangindent 20pt}

\pagestyle{empty}


\begin{center}
\bigskip
\vspace{.2in}
{\bf Do SNe Ia Provide Direct Evidence for Past Deceleration of the Universe?}\\

\bigskip

\vspace{.2in}
Michael S. Turner$^{1,2}$ and Adam G. Riess$^3$\\

\vspace{.2in}

{\it $^1$Departments of Astronomy \& Astrophysics and of Physics\\
Enrico Fermi Institute, The University of Chicago, Chicago, IL~~60637-1433}\\

\vspace{0.1in}
{\it $^2$NASA/Fermilab Astrophysics Center\\
Fermi National Accelerator Laboratory, Batavia, IL~~60510-0500}\\

\vspace{0.1in}
{\it $^3$Space Telescope Science Institute\\
3700 San Martin Drive, Baltimore, MD~~21218}\\

\end{center}

\vspace{.3in}
\centerline{\bf ABSTRACT}
\bigskip
\medskip

Observations of SN 1997ff at $z\sim1.7$ favor the accelerating
Universe interpretation of the high-redshift type Ia supernova
data over simple models of intergalactic dust
or SN luminosity evolution.  Taken at face-value,
they provide direct lines of evidence
that the Universe was decelerating in the past,
an expected but untested feature of the current
cosmological model.
We show that the strength of this conclusion
depends upon the nature of the dark energy
causing the present acceleration.
Only for a cosmological constant is the SNe evidence
definitive.  Using a new test which is independent of the contents of
the Universe, we show that the SN data favor recent
acceleration ($z < 0.5$) and past deceleration ($z>0.5$).

\newpage
\pagestyle{plain}
\setcounter{page}{1}

\section{Introduction}

The discovery in 1998 that the Universe is speeding up and not
slowing down (Riess et al, 1998; Perlmutter
et al, 1999) was a startling ``u-turn'' in the quest to finally
pin down the second
of Sandage's two numbers ($H_0$ and $q_0$; Sandage 1961, 1988):  The deceleration
parameter, $q_0$, is negative.
Observations of SN 1997ff, a type Ia supernova (SN Ia)
at $z \sim 1.7$ (Riess et al 2001; Gilliland, Nugent, \& Phillips 1999),
 and one or two more at $z \sim 1.2$ (Tonry et al. 2001; Aldering et al. 2001) disfavor
the two simple alternates to an accelerating Universe --
intergalactic dust and SN luminosity evolution (see e.g.,
Drell, Loredo and Wasserman 2000; Aguire 1999a,b; Riess 2000) -- and bolster
the case for the accelerating Universe hypothesis.
CMB measurements that indicate the Universe is flat
(Jaffe et al, 2001; Pryke et al, 2001; Netterfield
et al, 2001) and the failure of the matter density (by a factor of
three) to account for the critical density provides supporting
evidence for an additional component of energy density with
repulsive gravity (see e.g., Turner, 2000a).

There are very good theoretical reasons to believe
that accelerated expansion is a recent phenomenon.
A long matter-dominated phase is needed for the observed structure
to develop from the small density inhomogeneities revealed
by COBE and other anisotropy measurements made since
(Turner \& White, 1997).  Further, the success
of big-bang nucleosynthesis in predicting the light-element
abundances (see e.g., Burles et al, 2001),
and most recently the stunning confirmation of
the BBN baryon density by CMB anisotropy measurements
(Pryke et al, 2001; Netterfield et al, 2001),
is strong evidence that the Universe was radiation
dominated when it was seconds old.  The gravity of both matter and
radiation is attractive, leading to a strong theoretical
prejudice for an early
decelerating phase, lasting from at least as early
as 1 sec to until a few billion years ago.  However, this required feature of the
current cosmological model remains largely untested.

The purpose of our {\em Letter} is to assess
the strength of the {\em direct} empirical evidence
from type Ia SNe for slowing expansion in
the past.

\section{Preliminaries}

With good precedent (Robertson 1955; Hoyle \& Sandage 1956), 
we introduce a generalized, epoch-dependent deceleration
parameter\footnote{As can be seen from this equation, accelerated
growth of the cosmic-scale factor, i.e., $\ddot R > 0$, corresponds
to $q<0$; accelerated expansion rate, $\dot H > 0$, corresponds to $q<-1$.}
\begin{equation}
q(z) \equiv (-\ddot R /R)/H^2(z) = {d H^{-1}(z) \over dt} -1
\label{eq:q(z)}
\end{equation}
where $R(t)$ is the cosmic-scale factor (normalized to
be unity today), $H(z)=\dot R/R$ is the expansion rate
and $q_0 = q(z=0)$.  Just like $q_0$, for a matter-dominated,
flat Universe, $q(z)=0.5$ and for a vacuum-energy dominated,
flat Universe $q(z)=-1.0$.

A useful measurement of the change in the current expansion rate during
the span of a human time interval is far beyond
the precision attainable by known cosmological probes (but see Loeb 1998).  However,
measurements of distant supernovae can probe the expansion
{\it history} by determining luminosity distances, which
in turn are related to the integral of the inverse of the
preceding expansion rate.  For a flat Universe
\begin{equation}
d_L(z) = c(1+z)\int_0^z\, {du\over H(u)}\,,
\label{eq:dL}
\end{equation}
and more generally,
\begin{equation}
d_L(z) = c(1+z)|1-\Omega_0|^{-1/2} H_0^{-1}{\rm S}
	\left[|1-\Omega_0|^{1/2}H_0 \int_0^z dz/H(z)\right]
\end{equation}
where S$(x) = \sin (x)$ ($\Omega_0 > 1$), $\sinh (x)$
($\Omega_0 < 1$), and $x$ ($\Omega_0 =1$).  The quantities
$H_0$ and $\Omega_0$ refer to the current
($z=0$) Hubble constant and the sum of today's
energy densities in units of the critical density
($\rho_{\rm crit}=3H_0^2/8\pi G$), respectively.  The
comoving distance to an object at redshift $z$ is always
$r(z) = d_L/(1+z)$.

Equation \ref{eq:dL} can be rewritten in terms of the epoch-dependent
deceleration parameter of Eq (1):
\begin{equation}
d_L =c(1+z)\int_0^z {du\over H(u)} =
c(1+z)H_0^{-1} \int_0^z\,du \exp \left[-\int_0^u\,[1+q(v)]d\ln (1+v) \right]
\label{eq:dLq}
\end{equation}
again for a flat Universe, though easily generalized as above.
It is worth noting that only the assumption of the Robertson -- Walker
metric underlies Eqs \ref{eq:q(z)} - \ref{eq:dLq}.  Said another
way, deceleration/acceleration can be probed without assuming
the validity of general relativity or without providing a manifest 
of the contents of the Universe.  In the absence
of the Friedmann equation of general relativity to relate
the curvature radius to the matter/energy content, $cH_0^{-1}/|1-\Omega_0|^{1/2}$
is replaced by the spatial curvature radius.

Since supernovae measurements determine luminosity distances,
they cannot directly measure the
instantaneous expansion rate or deceleration rate.
(Number counts of standard objects, which depend upon
$r^2(z)/H(z)$, together with SNe, could in principle
determine $H(z)$ directly; see Huterer \& Turner, 2000.)
To use SNe to probe the expansion history,
one must make assumptions about the evolution of $H(z)$ or
$q(z)$; in turn, we shall take both approaches.

\section{Simple dark-energy models:  $\Lambda$ and const $w_X$}

While surprising, accelerated expansion can be accommodated within the
framework of the standard FRW cosmological model.
According to general relativity, the source of gravity is proportional
to $(\rho + 3p)$, stress-energy with large, negative pressure,
$p_X < -\rho_X/3$, has repulsive gravity.  In the absence of
an established cause for cosmic acceleration, the causative
agent has been referred to as ``dark energy.''  (In relativity
theory, any substance with pressure comparable in magnitude to
its energy density is relativistic -- more energy-like than
matter-like, and hence the name dark energy.)

The simplest possibility for dark energy
is the energy of the quantum vacuum (mathematically
equivalent to a cosmological
constant), for which $p_{\rm vac} = -\rho_{\rm vac}$.  However,
the natural scale for vacuum energy is at least 55 orders-of-magnitude
too large to allow the formation of structure by gravitational
instabilities in the early Universe (see e.g., Weinberg, 1989 or
Carroll, 2000).  The implausibility of reducing this by
precisely 54 orders-of-magnitude, suggests to some the
existence of an unrecognized symmetry that requires the energy of the
quantum vacuum to be precisely zero.  If this is so,
then some other source for the accelerated expansion is required.
Theorists have put forth a plethora of examples, from a rolling
scalar field (a mini episode of inflation, often called quintessence;
see e.g., Peebles \& Ratra 1988 or Caldwell, Dave, \& Steinhardt 1998)
to the influence of hidden additional space dimensions (Deffayet,
Dvali, \& Gabadadze 2001; for reviews see
Carroll, 2000; Turner, 2000b; or Sahni and Starobinskii, 2001).

For most purposes, the dark energy can be considered to be a smooth
component characterized by its equation-of-state, $w_X \equiv
p_X / \rho_X$, which may be a function of time (Turner \& White, 1997).
Doing so, and allowing for the fact that $w_X$ may vary with time,
the Friedmann equation for the expansion rate can be written as
\begin{eqnarray}
H^2(z) &=& H_0^2 [ \Omega_M(1+z)^3 + \Omega_X \exp [-3\int_0^z 
	(1+w_X(u))d\ln (1+u)] \nonumber \\
        & & \qquad + \Omega_R(1+z)^4 + (1-\Omega_0)(1+z)^2 ]
\end{eqnarray}
where $\Omega_i$ refers to the present fraction of critical density
in matter ($i=M$), in dark energy ($i=X$), and in radiation ($i=R$).
The final term (curvature term) vanishes for a flat Universe; the radiation
term, $\Omega_R \sim 10^{-4}$, is negligible for $1+z \ll 10^3$.
Neglecting radiation, the generalized deceleration parameter of Eq (1)
can then be written as
\begin{equation}
q(z)  =  {\Omega_0\over 2}+{3\over 2}w_X(z)\Omega_X(z)
\end{equation}

Specializing to a flat Universe, as indicated by recent CMB
anisotropy measurements which determine
$\Omega_0 = 1\pm 0.04$ (Jaffe et al, 2000; Pryke et al, 2001;
Netterfield et al, 2001), and constant $w_X$, these expressions become
\begin{eqnarray}
H^2(z) & = & H_0^2[\Omega_M(1+z)^3 + \Omega_X (1+z)^{3(1+w_X)}] \\
q(z) & = & {1\over 2} \left[ {1+(\Omega_X/\Omega_M)
(1+3w_X)(1+z)^{3w_X} \over 1+(\Omega_X/\Omega_M)(1+z)^{3w_X}}\right]
\end{eqnarray}
From this it follows that the redshift of transition from
deceleration to acceleration ($\equiv z_{q=0}$) is
\begin{eqnarray}
1+z_{q=0} & = & \left[ (1+ 3w_X)(\Omega_M-1)/\Omega_M \right]^{-1/3w_X} \nonumber \\
& = & \left[ 2\Omega_\Lambda /\Omega_M \right]^{1/3}
\label{eq:zlambda}
\end{eqnarray}
where the second equation is for vacuum energy (i.e., $w_X=-1$).

To begin, let us assume that the dark energy
is simply the energy of the quantum vacuum ($w_X=-1$).
If this is the case, the Universe must have been decelerating
in the past, for $z>z_{q=0}$ (provided that $\Omega_M>0$).
Still, we may ask, have we seen direct evidence of that deceleration (yet)?

The current SN Ia sample (see Riess et al. 1998; 
Perlmutter et al. 1999; Tonry et al. 2001) provides
measurements of the luminosity distance out to a redshift of
$z \sim 1.7$ with the extreme redshift provided by SN 1997ff (Riess et al. 2001).
From Eq (11) it follows that accelerated
expansion throughout the interval sampled by the SNe
is equivalent to $\Omega_M < 0.09$.
Using the data employed by Riess et al. (2001), we have constructed
the a posteriori probability density for $\Omega_M$.
The null hypothesis (i.e., $\Omega_M < 0.09$)
is rejected with greater than 99.9\% confidence.
To be specific, the 99\% confidence interval is
\begin{equation}
0.14 <\Omega_M< 0.60
\end{equation}
Further, SN 1997ff alone is inconsistent with $\Omega_M < 0.09$ at about
the 99\% confidence level.

Without SN 1997ff there is little direct evidence for past deceleration.  The next highest
redshift supernova in the sample used by Riess et al. (2001) 
was at $z\simeq 1$ (SN 1997ck; Garnavich et al. 1998a). 
In order to conclude that this supernova had directly probed the deceleration, would 
require constraining
$\Omega_M > 0.2$, cf. Eq (11).  The 99\% confidence interval for the SNe
used in Riess et al (2001), excluding SN 1997ff, is $0.11 < \Omega_M < 0.58 $.

Now, consider dark-energy models with constant equation-of-state
$w_X$ (or approximately constant for $z<1.7$).
Assuming once again a flat Universe, this leaves two cosmological
parameters:  $w_X$ and $\Omega_M$.  For constant $w_X$ models, the
Universe always has a decelerating phase at high-redshift,
cf. Eq (10).  Provided the matter density is sufficient,
the epoch of transition from acceleration to deceleration
occurs at $z_{q=0} < 1.7$; specifically, if
\begin{equation}
\Omega_M > {1 \over 1 - (2.7)^{-3w_X}/(1+3w_X)}\,.
\end{equation}
The region in the $w_X$ -- $\Omega_M$ plane where the transition
from acceleration to deceleration occurs at $z<1.7$ is shown in Fig.~1.

\begin{figure}
\centerline{\psfig{figure=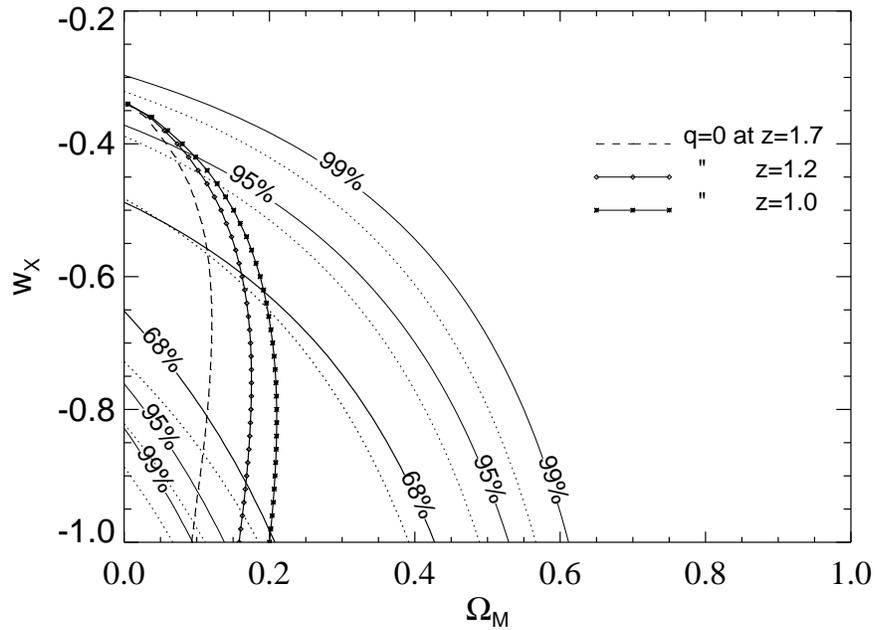,width=5in,angle=90}}
\caption{
The solid contours show the regions of probability
indicated; the dotted contours show the same, but without
SN 1997ff.  The three horizon curves delineate the mass
density required so that the transition from acceleration
to deceleration occurs at $z=1.7,1.2,1.0$ (see text).
Note:  while indicative of the $w_X=-1$ case considered
previously, the contours cannot be directly used to
infer the confidence range for $\Omega_M$ for $w_X = -1$.
}
\end{figure}

Also shown in Fig.~1 are confidence contours for the data employed
by Riess et al (2001),
computed with and without SN1997ff (see also
Garnavich et al. 1998b; Perlmutter et al. 1999).
As can be seen, SN1997ff significantly
increases confidence that the transition from acceleration
to deceleration occurred within the redshift interval sampled
by the supernovae.  Even so, only for $w_X$ near $-1$, is
$z_{q=0}$ constrained to less than 1.7 with high confidence.

For $w_X > -0.9$, SNe alone do not provide
significant, direct evidence for past deceleration.
However, with an external and reasonable constraint based upon dynamical
measurements of $\Omega_M > 0.12$ (e.g., see Primack, 2000; or Turner, 2001),
deceleration is guaranteed for $z < 1.7$.

SNe Ia at the next highest redshift known, $z=1.2$,
(e.g., SN 1999fv; Tonry et al. 2001, SN 1998eq;
Aldering et al. 1999) or $z=1.0$ (e.g., SN 1997ck; 
Garnavich et al. 1998a; SN 1999fk; Tonry et al. 2001) 
are likely too near to provide a direct requirement for 
deceleration at the time of their explosion due to the 
reduced leverage at these redshifts (see Figure 1).
Even indirect evidence of deceleration is marginal at 
these redshifts requiring $\Omega_M > 0.2$.

\section{A new, model-independent test for deceleration}

As discussed in \S 2, luminosity distance can be written in
terms of the epoch-dependent deceleration parameter, $q(z)$,
with only the assumption of the Robertson-Walker
metric, cf. Eq \ref{eq:dLq}.  Proceeding from this equation,
one can test for past deceleration in the most general way.

As a null hypothesis, suppose that the Universe
never decelerated across the redshift
interval sampled by the current set of supernovae:
that is, $q\le 0$ for $0<z<1.7$.  Using the fact that
$-1 < -(1+q)$ if $q \le 0$, it then follows that
\begin{equation}
d_L(z) \ge cH_0^{-1}(1+z)\ln (1+z)
\end{equation}
The logic of the inequality is clear:  in a
universe that is always coasting or accelerating,
objects of a given redshift are farther away (and fainter)
than in a universe that at some time has decelerated.  (Note:
the equality applies for a flat, eternally coasting universe,
which can be achieved with $\Omega_X =1$ and $w_X=-1/3$.)

Using the measurements for SN1997ff, this inequality reads
$$
c^{-1}H_0d_L(z=1.7) = 2.4\pm 0.4 > 2.7
$$
The null hypothesis (Universe has never decelerated between
$z=0$ and $z=1.7$) is violated, though
with little significance.  However general and simple,
this analysis does not take into account the evidence for recent
acceleration and thus dilutes the possible evidence for past deceleration.

We can increase our resolution to past episodes of deceleration
by considering a sharper, two-epoch model
that allows for the possibility of
a change in the deceleration parameter
\begin{eqnarray}
q(z) & = & q_1\qquad {\rm for\ }z<z_1 \\
     & = & q_2\qquad {\rm for\ }z>z_1 \nonumber
\end{eqnarray}
The motivation for this ansatz is to test for what theory and data suggest:
early deceleration (i.e., $q_2 > 0$) followed
by recent acceleration (i.e., $q_1 < 0$), without specializing
to a particular dark-energy model, or even assuming that
the Friedmann equation describes $H(z)$.  Physically, the parameters
$q_1$ and $q_2$ correspond to average deceleration parameters
for redshifts less than $z_1$ and greater than $z_1$
respectively.

For this two-parameter model, it is straightforward to
obtain the luminosity distance:
\begin{eqnarray}
d_L & = & (c/H_0)(1+z){\left[ 1 - (1+z)^{-q_1}\right] \over q_1}
        \qquad z<z_1 \\
         & = &  (c/H_0)(1+z)\left[{\left[1 - (1+z_1)^{-q_1}\right]\over q_1}
         +{ (1+z_1)^{q_2-q_1}\left[ (1+z_1)^{-q_2} - (z+z)^{-q_2}\right]
         \over q_2}\right] \qquad z>z_1 \nonumber
\end{eqnarray}

 The transition redshift
$z_1$ is arbitrary.  However, due to the limited sampling of the redshift interval
 provided by the SNe and our interest in resolving the
 behavior in {\it both} regions, we selected
values of $z_1$ near $z_1 \sim 0.5$.

\begin{figure}
\centerline{\psfig{figure=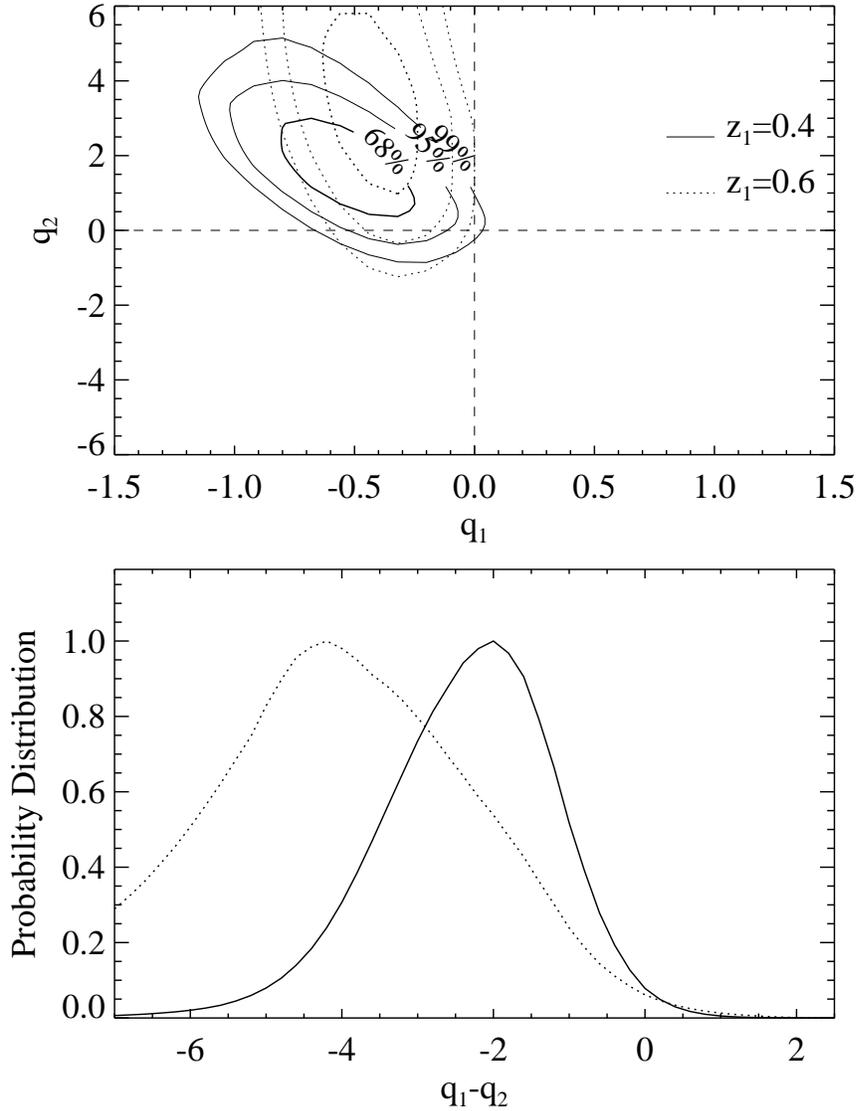,width=5in}}
\caption{
(a)  Probability contours in the $q_1$ -- $q_2$
plane; solid curves are for $z_1 = 0.4$ and dotted for $z_1 = 0.6$.
The four quadrants correspond to the four different acceleration
histories.  With better than 90\% confidence
the SN data prefer recent acceleration ($q_1 < 0$) and
past deceleration ($q_2 > 0$).  (b) Probability distribution for
$q_1-q_2$; solid curve for $z_1=0.4$ and dotted curve
for $z_1=0.6$.  The SN data strongly indicate lessening acceleration
with redshift -- the assertion of attractive gravity at around
$z\sim 0.5$.
}
\end{figure}

In panel (a) of Fig.~2 we display confidence contours in
the $q_1$ -- $q_2$ plane, for $z_1 = 0.4$ and $z_1 = 0.6$ using the current SN sample.
The four quadrants of the plot represent the four
histories for cosmic expansion; past and present acceleration or
deceleration, or a transition in the sign of $q_i$ across $z_1$.

Not surprisingly, it is possible to reject both right quadrants, for which
$q_1> 0$, with very high confidence.  Recent acceleration is a
robust feature of the SN data.  For the
remaining two quadrants, the upper left quadrant ($q_2 > 0$)
indicating past deceleration is preferred by the data, but we cannot exclude
past acceleration ($q_2 < 0$) with great confidence
($\sim 90\%$ confidence).

We have constructed the contours for the current SN Ia sample
excluding SN 1997ff; while the evidence for recent acceleration remains equally
significant, the evidence for past deceleration is much less
significant.  Although SN 1997ff provides the
greatest leverage for any single SN for this test, more SNe Ia at
$z>1$ are needed to sharpen this model-independent,
direct test for past deceleration.

Finally, in panel (b) of Fig.~2 we show the a posteriori
probability distribution for $q_1 - q_2$.  The SNe data
strongly prefer an increase in $q(z)$ with increasing redshift.
This is a strong, model independent indication for a change in the deceleration rate
with time in the sense of moving from recent acceleration to
past deceleration.  Said another way, we see direct evidence for the
assertion of attractive gravity in the past.

\section{Concluding remarks}

The absence of an early, decelerating phase would be a
much bigger surprise than the discovery that the Universe is
accelerating today.  It would be essentially impossible to
reconcile with the standard hot big-bang cosmology.
In addition to providing strong support for the accelerating Universe
interpretation of high-redshift SNe Ia, SN 1997ff
at $z \sim 1.7$ provides direct evidence for an early phase 
of slowing expansion {\it if} the dark energy is a 
cosmological constant (Riess et al. 2001).
However, because supernova observations do not directly 
measure changes in the expansion
rate, a model for $H(z)$ or $q(z)$ is needed to perform
a more robust test for past deceleration.
The former requires assumptions (or a deeper understanding) about 
the nature of dark
energy responsible for the recent speed up while 
the latter requires more SNe Ia at $z > 1$.

In our analysis we have employed the measurement of SN 1997ff by
Riess et al. (2001) at ``face-value''.  Riess et al. (2001) discuss 
a number of possible contaminants to this measurement which, 
if manifested, could significantly reduce the cosmological utility of 
this supernova.  For example, host extinction could make the SN appear 
dimmer or foreground lensing (Lewis \& Ibata 2001; Moertsell, 
Gunnarsson, \& Goobar 2001) could make the SN brighter.
However, a face-value treatment of this SN is plausible as 
significant contaminantion, while possible, appears to be unlikely.  
The star formation history of the red, elliptical host of SN 1997ff 
suggests that substantial foreground extinction of the supernova is 
not likely.  Likewise,
due to the lack of apparent shear of the host galaxy, the 
simplest interpretation is that the SN is not greatly magnified by 
the nearest foregrounds (as opposed to more complex scenarios in 
which the SN is highly magnified {\it and} a corresponding tangential shear 
of the host counteracts an intrinsic, radial elongation of the host; Riess et al. 2001).

In summary, we have shown that for a flat Universe the current supernova data:
\begin{enumerate}

\item provide strong, direct evidence of past deceleration
{\it if} it is assumed that the dark energy is vacuum energy
(cosmological constant).

\item alone do not provide direct evidence of past deceleration unless
the dark-energy equation-of-state $w_X$
is close to $-1$.  However, using dynamical measurements of
the amount of matter, deceleration can be indirectly inferred for
the redshift range of the SN sample (if $\Omega_M > 0.12$).

\item  without recourse to a specific model of the contents
of the Universe,  favor deceleration at $z > 0.5$ with $\sim$90\% confidence.
An even stronger statement is that the SN data favor increasing $q(z)$
with increasing redshift, a sign of the assertion of attractive
gravity in the past.
\end{enumerate}

What then can make the SN Ia evidence for a decelerating phase in the
past stronger?  Additional high-redshift ($z> 1$) supernovae
would strengthen both the model-independent
and the $w_X$ -- $\Omega_M$ analysis.  Interestingly enough,
very-low redshift supernovae also have significant leverage
by reducing the uncertainty in the
contemporary expansion rate (although uncertainty in the
zeropoint calibration of SNe Ia does not affect the analysis).
Specifically, if we had fixed
the Hubble constant, the $q_1$-$q_2$
analysis would have implied a past decelerating phase with
greater than 95\% confidence and the 68\% confidence contours in the $w_X$ --
$\Omega_M$ plane would have closed at $\Omega_M > 0$.
  Fortunately, systematic programs are underway
to garner many more SNe Ia at both low and high redshifts.

High-redshift SNe Ia fill a unique niche in the toolbox of observational cosmology.
As demonstrated here, they can provide a direct test of past deceleration, a salient
and testable prediction of our current cosmological paradigm.  In addition, they have
great potential to unlock the mystery of the nature of the dark energy.

This work was supported in part by the DOE (at Chicago) and by the
NASA (at Fermilab and STScI).

\end{document}